IRAD 2018 JOURNAL PAPER# Reconfigurable Electromagnetic Structures: X-Band and Beyond

Kenneth W. Allen and Joshua M. Kovitz
**Abstract.** Developing reconfigurable millimeter-wave (mmWave) antennas and devices is an outstanding challenge, with switch technologies being a primary impediment. Recently, it has been shown that vanadium dioxide ($VO_2$), a thermochromic material whose resistance changes with temperature, could provide a path forward in developing reconfigurable mmWave devices. As an initial step towards this vision, we investigate the integration of $VO_2$ switches in reconfigurable components at 15 GHz. In particular, a frequency reconfigurable antenna and a reconfigurable phase shifter are shown. The low loss and minimal parasitics of $VO_2$ technology have the potential to enable devices at 15 GHz and beyond.

**Introduction.** Reconfigurable electromagnetic structures, where switches alter the current distribution to provide functionality, have drawn significant attention in the research community for the last two decades. Reconfigurability has become integral as the needs for multifunctional radar and communications systems continue to grow, consequently, increasing demands on each component in the signal chain of the system. These demands have also led to RF congestion at microwave frequencies < 10 GHz. Engineers in the telecommunications, defense, and other commercial industries have set their sights on developing mmWave systems [1]-[4] in order to utilize the less congested mmWave spectrum. It is also anticipated that these systems will be versatile and multifunctional. Integrating switches into microwave devices is common practice, primarily because switches offer reliability and high dynamic range. Unfortunately, a limited number of switch technologies have shown promise at mmWave frequencies [18]-[22]. Recent advancements with phase change material (PCM) switches, however,

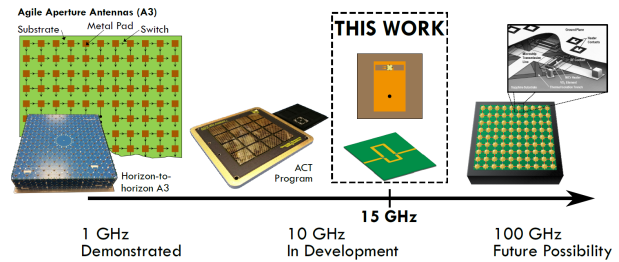

**Figure 1.** The path forward towards integrating $VO_2$ switches into mmWave devices. Inset $VO_2$ switch figure reproduced from [7]. [11]

suggest that mmWave reconfiguration might be close to realization [11].

Recently, it has been demonstrated that PCM switches using vanadium dioxide ($VO_2$) can achieve encouraging performance within the mmWave bands and beyond [5]-[12]. These $VO_2$ switches exploit the metal-insulator transition of $VO_2$ which is controlled by the material's temperature, and the switch can be thermally driven using an integrated microheater. These findings are timely with recent initiatives within the Georgia Tech Research Institute (GTRI) to further develop agile aperture antennas (A3) at higher frequencies. GTRI has a long and successful history of developing highly reconfigurable antennas using a switch/pixel configuration at lower microwave frequencies (<6 GHz) with traditional switches [15]. More recently, the research has pushed to higher frequencies, e.g. X-band [16]. Our vision for future developments is to package similar architectures at mmWave frequencies, as illustrated in Figure 1. The aim is to develop mmWave A3 platforms that can be integrated with high performance switches monolithically, and $VO_2$ switches are among the few technologies capable of performing at higher frequencies. In addition to these versatile apertures with a plethora of operating states, we will consider leveraging fragmented aperture-style antennas [25]-[28] and combiners [28] with the addition of strategically placed switches for a limited number of states tuned for specific operations. This IRAD paper presents an initial "stepping-stone" evaluation around 15 GHz, based on [10]-[11], where we study several reconfigurable components and antennas. This $VO_2$ switch technology has recently been developed by Teledyne, and extensive characterization has been presented in [6]-[9].

**Technical Approach.** In this phase of the IRAD work, our goal was to develop detailed simulations models of the switch, develop interconnects between a PCB and the switch, and prototype switch fixtures for validation of the models. The foundation of all this work was the development of a detailed switch model that would be used to model the switch in both reconfigurable circuits and reconfigurable antenna settings.

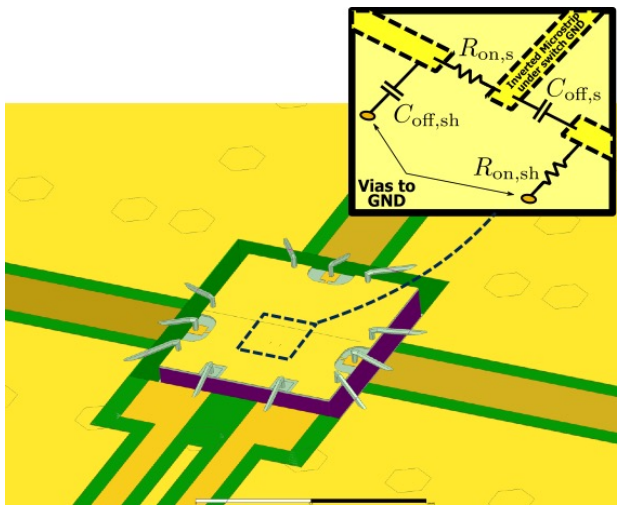

**Figure 2. Three-dimensional CAD model of the $VO_2$ switch in HFSS. As a narrowband simplification, we chose to use an approximate circuit model to incorporate the measured performance at 15 GHz [8].**

Modeling the switch within a full-wave CAD modeling software (HFSS) is an important step relevant to switch integration. There are a variety of techniques that have been used in the past (reference [20] summarizes several for MEMS switches). In this work, we opted for a hybrid circuit/full CAD model of the switches as a balance between simulation time and accuracy. Figure 2 illustrates the main features of this $VO_2$ switch model, where the stackup of the switch is also depicted. The switch buildup is identical to that in [6]-[8], where an inverted microstrip is coupled to our external PCB artwork through $70\mu m \times 70\mu m$ pads and $25\mu m$ gold wirebonds.

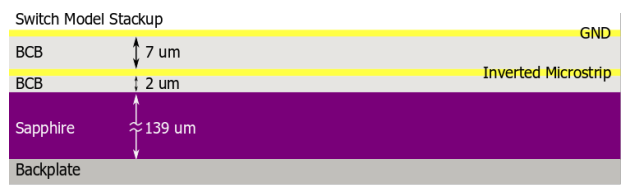

**Figure 3. Representative stackup of the switch model**

We reduce the switch internals to a simple reflective style SPDT switch where the OFF branch (RF2) is connected to RFC by a series capacitance $C_{off,s}$ and the ON branch (RF1) is connected to RFC by a series resistance $R_{on,s}$. The reflective shunt impedances $R_{on,sh}$ and $C_{off,sh}$ are directly connected to GND. This greatly simplifies model placement in HFSS and also models potential effects on radiation patterns from the switch internal structures. It should be noted that this simplification can only support narrowband applications, where switch impedances are reasonably well-behaved over the frequency band. The values of $R_{on}$ and $C_{off}$ depend on the VO2 channel width and length. For our model, we used the values based on curve-fitting S-parameters from the L1W4 and L2W4 switch models as done in [8]-[9].

**Technical Results.** In total, we developed three test structures that would prove the usefulness of the switch at both the component level and a system level. The first item was a test fixture that would allow us to characterize and develop an interface with the switch using wirebonds. The second developed test fixture was a phase shifter design that uses two VO₂ SPDT switches, and the last design was a frequency reconfigurable patch antenna.

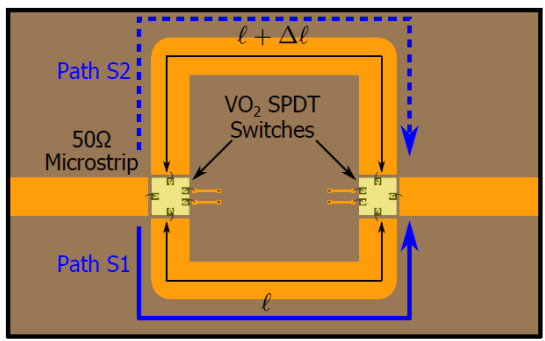

Figure 4. Low loss 1-bit phase shifter using VO₂ SPDT switches for reconfigurable feed networks. [11]

The switch measurement fixture was an important article to develop to move forward with further integration concepts. If built, this structure would also allow us to characterize both the switch and interconnect structure developed within the scope of this IRAD. Figure 2 shows a magnified view of the simulated model used for verification and interconnect development. In essence, three CPW lines with very small trace-to-ground gaps connect to the switch RF interface. The geometry of these traces had to be carefully tuned to achieve working performance. It was also found that using the wirebond configuration shown in Figure 2, where one wirebond connects the switch RF trace to the CPW trace and two nearly parallel wirebonds connect the switch ground to CPW ground, provided good performance. Once satisfactory results were obtained for this switch and its model, i.e. $S_{11} < -25$ dB from $14 - 16$ GHz, then this model was then used to further develop the phase shifter and reconfigurable antenna.

Another interesting test is to develop a large scale, low-loss phase shifter as shown in Figure 4. The concept uses variable length delay lines to achieve a given phase delay of 90° at 15 GHz, as illustrated in the phase plot of Figure 5. Such a concept has applications in low-loss beamformers and reconfigurable feed networks for polarization agility [23]-[24]. This circuit also provides an accurate testbed for switch modeling and calibration campaigns.

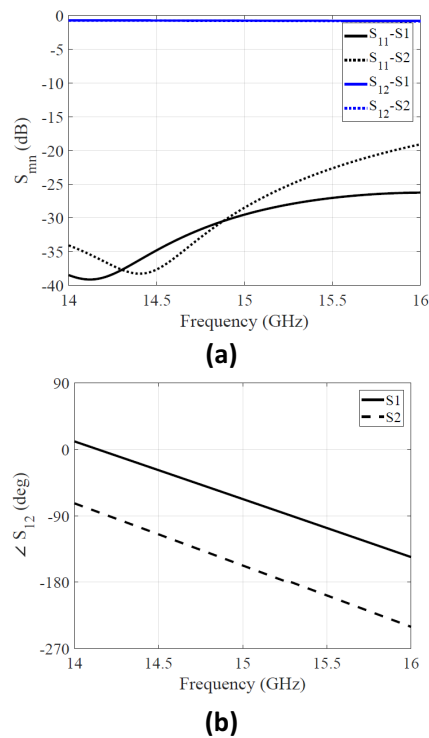

Figure 5. (a) Modeled S-parameters of the 1-bit phase shifter using VO₂ switches. (b) Phase difference between the two different states S1 and S2. [11]

The phase shifter was developed on a Rogers RO4003C 0.008" substrate. To integrate the switch, a portion of the substrate was excised to generate rectangular holes and a supporting backplate is assumed (magnified model in Figure 2). The 3D model of the VO₂ switch was completed in HFSS, while the overall phase shifter was cascaded and simulated in ADS. This included losses for the microstrip lines as well as the loss of the switches, and S-parameter performance is illustrated in Figure 5. For state S1 (the shorter path), the insertion loss averaged roughly 1.09 dB, while the state S2 (the longer path) had an average insertion loss of 1.15 dB. Comparing the case of ideal transmission lines, the insertion loss coming from the switches is roughly 0.94 dB, providing around 0.47 dB insertion loss per switch at 15 GHz, which agrees with measurements done in [9].

Frequency agility is a desirable capability for antennas. An alternative path to developing wideband antennas, which can provide additional filtering at the aperture level, is to integrate switches in narrowband antennas to support different radios and functionality. Frequency reconfiguration can be accomplished in many ways, but it is most often done by tuning the resonance behavior of a given narrowband antenna. A frequency reconfigurable patch antenna is shown in Figure 6, where a $VO_2$ switch is placed within a slot cut into a typical rectangular patch (similar to [29]). When both switches are in the ON state, the patch appears like a longer rectangular patch with a slot cut near the edge and provides a lower resonant frequency. When both switches are in the OFF state, the patch appears shorter and results in a higher resonant frequency.

The $VO_2$ switch model discussed in the previous section was incorporated into our HFSS model to demonstrate reconfigurability and evaluate effects of the switch. The dielectric substrate used in this design was Rogers RT Duroid 5880LZ, having 0.040" thickness. The ground plane and metallized areas of the patch are modeled as copper sheets with surface roughness values given by Rogers and incorporated using the Groisse model. The bias pads are connected through a via that bypasses the ground plane (not DC connected). This bias scheme can detune the resonance; however, full-wave optimization can compensate for these slight deviations from an ideal model. The performance is summarized in Figure 7, where impedance matching clearly demonstrates a shift in the frequency resonance. A representative radiation pattern in the off-state resonant frequency is also shown in Figure 7, and patterns in the on-state also exhibited similar features. The predicted radiation efficiency (excluding mismatch) of this antenna is -0.36 dB and -0.72 dB at resonance for the OFF (15.6GHz) and ON (15.0GHz) states, respectively. This simulation included dielectric and conductor loss (including surface roughness). If the switch had been ideal with $R_{on} = 0\ \Omega$, then the efficiency for the ON state increases to -0.43dB, indicating that only 0.29 dB is lost to the switch.

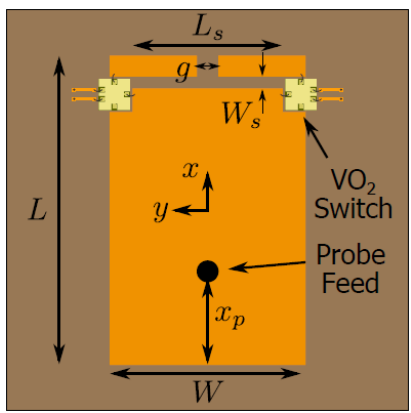

Figure 6. A frequency reconfigurable patch antenna using $VO_2$ switches. [11]

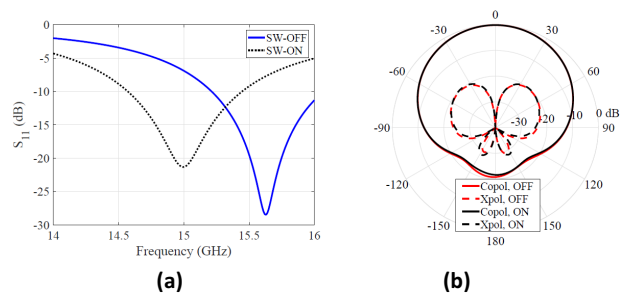

(a)   (b)

Figure 7. (a) Modeled $S_{11}$ performance of the frequency reconfigurable antenna. (b) H-plane (YZ) pattern cuts for the frequency reconfigurable antenna at each states' resonant frequency. [11]

At the end of this work, we also attempted to build some prototypes of the structures developed within the IRAD. The switch measurement fixture was the first priority to validate our simulated models shown previously. This involved a number of steps towards the integration of these switches into a PCB. An important fabrication decision was the development of a hole/cavity in the PCB so that the switch could sit at nearly the same plane as the other CPW traces. Pictures of the constructed switch measurement fixture and an integrated switch are shown in Figure 8, where three CPW traces connect to the switch under test. The partial prototype, as seen in Figure 8, was successful up to the integration stage, but we had trouble wirebonding to the switch. Our findings were that the gold layers had poor adhesion to the BCB layers that they were built on. Once a wirebond was placed, the force of pulling the wirebond over to the next pad was enough to delaminate the gold from the BCB, as illustrated in Figure 9. This was a setback that prohibited us from taking further measurements with the switches. Despite this setback, we plan to continue the investigation and work with Teledyne to explore other options for wirebonding these switches.

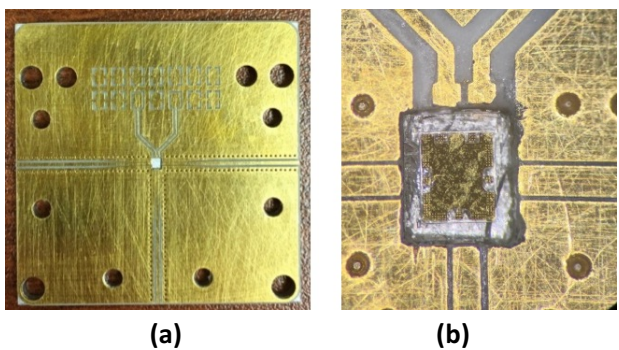

(a)          (b)

**Figure 8.** (a) Switch measurement fixture PCB. (b) Magnified view of a switch integrated into the PCB fixture for measurements (wirebonds not applied yet).

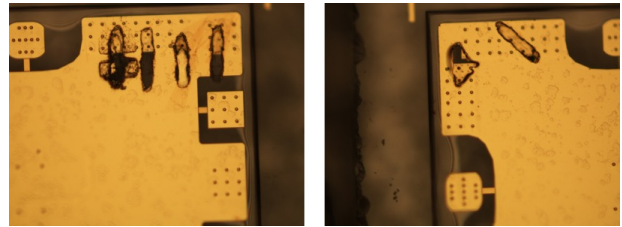

**Figure 9.** View of the wirebonds after attempting to place a wirebond. The gold metal layer underneath the top gold layer is present near the areas where the metal was delaminated.

**Conclusions.** This work highlights our recent modeling and prototype efforts with the grand vision towards integrating $VO_2$ switches for mmWave applications. Among many competing technologies, $VO_2$ switches show promising performance, especially at the mmWave and sub-mmWave frequencies. This work showed some simple examples and data suggesting the next generation of high-performance reconfigurable antennas and mmWave components. This initial first look focused on 15 GHz, but our goal is to integrate this technology into future mmWave systems beyond 40 GHz. The overall simulated performance and losses in the case of the reconfigurable antenna and reconfigurable phase shifter suggested a good performance was achieved.

**Acknowledgements.** The authors would like to thank Dr. Ryan S. Westafer and Dr. R. Todd Lee for stimulating discussions. This work was funded from the Georgia Tech Research Institute's independent research and development HIVE program under the direction of Mr. Benjamin Riley and Mrs. Kathy Harger; their guidance and support is greatly appreciated. The authors would like to thank Chris E. Hillman from Teledyne Scientific Company for donating the $VO_2$ switches and helpful discussions regarding models that have been used throughout this work.

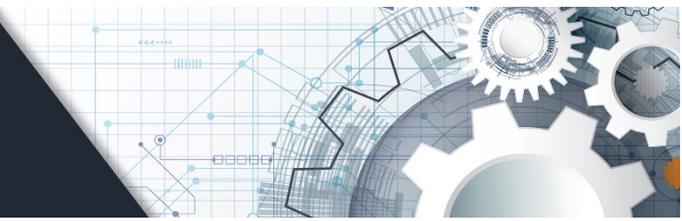

# IRAD 2018 JOURNAL PAPER

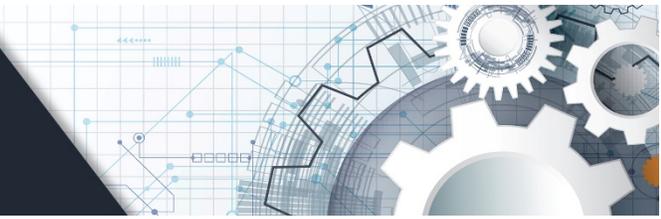